\documentclass[conference]{IEEEtran}
\usepackage{cite}
\usepackage{xcolor}

\ifCLASSINFOpdf
 \usepackage[pdftex]{graphicx}
\else
 \usepackage[dvips]{graphicx}
\fi

\newif\iflong
\longtrue
\iflong
	\newcommand{\vref}[1]{Section~\ref{#1}}
\else
	\newcommand{\vref}[1]{\cite{functions-mrfs-extended}}
\fi

\usepackage{amsmath,amssymb,amsthm}
\newtheorem{prop}{Proposition}
\newtheorem{lem}{Lemma}
\newtheorem{cor}{Corollary}
\theoremstyle{definition}
\newtheorem{defin}{Definition}
\newtheorem{example}{Example}
\newtheorem{problem}{Problem}
\newtheorem{remark}{Remark}

\usepackage{tikz}
\usepackage{verbatim}
\usetikzlibrary{arrows,shapes}
\tikzstyle{vertex}=[circle,fill=blue!25,minimum size=20pt,inner sep=0pt]
\tikzstyle{selected vertex} = [vertex, fill=red!24]
\tikzstyle{edge} = [draw,thick,-]
\tikzstyle{weight} = [font=\small]
\tikzstyle{selected edge} = [draw,line width=5pt,-,red!50]
\tikzstyle{ignored edge} = [draw,line width=5pt,-,black!20]

\newcounter{mparcnt}

\usepackage{balance}

\begin{document}

\title{On Functions of Markov Random Fields}

\author{
\IEEEauthorblockN{Bernhard C. Geiger}
\IEEEauthorblockA{Know-Center GmbH, 8010, Graz, Austria\\Email: geiger@ieee.org}
\and
\IEEEauthorblockN{Ali Al-Bashabsheh}
\IEEEauthorblockA{
%Beijing Advanced Institution on Big Data and Brain \\
%Computing, Beijing, China (Email: entropyali@gmail.com)
%
%Beijing Advanced Institution on Big Data and \\
%Brain Computing, Beihang University, Beijing, China \\
%Email: entropyali@gmail.com
Beijing Advanced Inst. on Big Data and Brain \\ 
Computing, Beijing, China  
(entropyali@gmail.com)
}
}

\maketitle

\begin{abstract}
We derive two sufficient conditions for a function of a Markov random field (MRF) on a given graph
to be a MRF on the same graph. The first condition is information-theoretic and parallels a recent
information-theoretic characterization of lumpability of Markov chains. The second condition, which
is easier to check, is based on the potential functions of the corresponding Gibbs field. We
illustrate our sufficient conditions at the hand of several examples and discuss implications for
practical applications of MRFs. As a side result, we give a partial characterization of functions of
MRFs that are information-preserving.
\end{abstract}

% Note that keywords are not normally used for peerreview papers.

\newcommand{\graph}{\mathcal{G}}
\newcommand{\sets}[1]{\mathcal{#1}}
\newcommand{\neigh}[1]{\mathcal{N}_{#1}}
\newcommand{\Prob}{\mathbb{P}}
\newcommand{\pmf}[1]{p_{#1}}
\newcommand{\mutinf}[1]{I(#1)}
\newcommand{\ent}[1]{H(#1)}
\newcommand{\entrate}[1]{\bar{H}(#1)}
\newcommand{\reals}{\mathbb{R}}
\newcommand{\e}[1]{\mathrm{e}^{#1}}

\newcommand{\vi}{i}
\newcommand{\vj}{j}
\newcommand{\cl}{\ell}
\newcommand{\cL}{L}
\newcommand{\perm}{v}

\section{Introduction}

Since the late 1950s, researchers have actively investigated properties of functions of Markov chains. In particular, considerable effort has been devoted to obtain sufficient and necessary
conditions for \emph{lumpability}, the rare scenario in which a function of a Markov chain has the
Markov
property~\cite{GurvitsLedoux_MarkovPropertyLinearAlgebraApproach,Kemeny_FMC,GeigerTemmel_kLump}.

In this work, we extend the concept of lumpability and its investigation  to Markov random fields (MRFs). Specifically, given a MRF
$X:=(X_1,\dots,X_N)$ on a graph $\graph$, we determine conditions for a set of functions
$\{g_1,\dots,g_N\}$ such that the transformation $Y:=(g_1(X_1),\dots,g_N(X_N))$ is a MRF on
$\graph$.
In other words, the problem we investigate asks the question under which functions an independence structure (i.e., a collections of independence statements) remains valid.

Aside from being an interesting problem in its own right, it is also practically
motivated from an inference perspective. Namely, multidimensional data $X$ is often modeled as a hidden
MRF, i.e., the data $X$ is hidden and can be inferred from some observed random variable 
$Z:=(Z_1,\dots,Z_N)$, where each $Z_i$ is conditionally independent of $X$ given $X_i$. 
In some scenarios, however, not $X$ is of interest but its transformation $Y$. For example, in
image processing, in which $\graph$ is a graph on a lattice with a distance-based neighborhood
structure and in which $X$ and $Z$ denote the true and observed pixel values, respectively, one may be interested in subsampling the image,
clustering regions of the image, or quantizing pixel values for the sake of identifying regions with
similar intensities. Transforming $X$ to $Y$ potentially creates additional or breaks existing dependencies, i.e., the graph $\graph_Y$ w.r.t.\ which $Y$ is a MRF is generally different from
$\graph$.
Rather than inferring $X$ from the observed $Z$ and subsequently computing $Y$ via the known
transformations, in this work, we are interested in scenarios where $Y$ is directly inferred from
$Z$. This is computationally tractable if $(Y,Z)$ turns out to be a hidden MRF itself. Among other things, this requires determining the graph $\graph_Y$ w.r.t.\ which $Y$ is a MRF.

The remainder of this paper can be summarized as follows. Section~\ref{sec:notation} introduces
notation and basic definitions, and Section~\ref{sec:formulation} formulates the problem and provides
some examples.
Section~\ref{sec:related} places the current work in context with previous results on stochastic
transformations of MRFs~\cite[Sec.~IV]{Perez_TIT} and subfields of
MRFs~\cite{Yeung_ISIT,Yeung_IT,Perez_TIT}.
Section~\ref{sec:sufficient} gives two sufficient conditions for
$Y$ to be a MRF on the same graph as $X$, i.e., for $\graph_Y=\graph$. 
The first condition is based on the characterization of MRFs via clique potentials, while the second 
is information-theoretic and resembles the information-theoretic characterization of Markov chain
lumpability~\cite[Th.~2]{GeigerTemmel_kLump}. 
As a side result, Section~\ref{sec:loss} presents necessary and sufficient conditions for the
transformation $Y$ to have the same information
content as $X$.
For the sake of readability,
\iflong\else and due to space limitations, \fi all proofs are in~\vref{sec:proofs}.

\section{Notation and Preliminaries}
\label{sec:notation}
Let $\graph:=(\sets{V},E)$ be an undirected graph with vertices $\sets{V}:=\{1,\dots,N\}$ and edges
$E\subseteq[\sets{V}]^2$, where $[A]^2$ is the set of two-element subsets of $A$.
We call $\graph$ complete if $E=[\sets{V}]^2$, chordal if every induced cycle of $\graph$ has length three, a tree if $\graph$ is connected and acyclic,
and a path if there is a permutation $\perm_1,\dots,\perm_N$ of the vertices such that
$E=\{\{\perm_i,\perm_{i+1}\},\ i=1,\dots,N-1\}$.
If $\{\vi,\vj\}\in E$, then
the vertices $\vi$ and $\vj$ are neighbors, and we use $\neigh{\vi}$ to denote the neighbors of $\vi$, i.e.,
\begin{equation}
    \neigh{\vi} := \{\vj \in\sets{V}\setminus \{\vi\}{:}\ \{\vi,\vj\} \in E\}.
\end{equation}
A set $C\subseteq\sets{V}$ is called a clique if it is a singleton or if
$[C]^2\subseteq E$. We use $\sets{C}$ to denote the set of cliques of $\graph$.

We denote random variables (RVs) by upper case letters, e.g., $X$, alphabets by calligraphic
letters, e.g., $\sets{X}$, and realizations by lower case letters, e.g., $x$. We assume that all our
RVs are defined on a common probability space $(\Omega,\sets{T},\Prob)$. Specifically, let
$X_{\vi}$ be a discrete RV with alphabet $\sets{X}_{\vi}$ that is associated with vertex $\vi\in\sets{V}$. For a set
$A\subseteq\sets{V}$, we write $X_A:=(X_{\vi},\ \vi \in A)$ and $\sets{X}_A:= \prod_{\vi \in A}
\sets{X}_{\vi}$.
We furthermore use the abbreviations $X:=X_{\sets{V}}$ and $X_{\not \vi}:=X_{\sets{V}\setminus
\{\vi\}}$, and similarly for the alphabets of these RVs. The RV $X_A$ is characterized by its
probability mass function (PMF)
\begin{equation}
  \pmf{X_A}(x_A) := \Prob(\{\omega\in\Omega{:}\ X_A(\omega)=x_A\}), \forall x_A\in\sets{X}_A.
\end{equation}

\begin{defin}\label{def:MRF}
	Let $\graph = (\sets{V}, E)$ be a graph and $X=(X_{\vi}, \vi \in \sets{V})$ be a RV with PMF $\pmf{X}$, then
$X$ is a \emph{Markov random field} (MRF) on $\graph$, abbreviated $X$ is a $(\graph,\pmf{X})$-MRF, if
 \begin{equation}
	 \label{eq:MRF}
	 \forall \vi \in\sets{V}{:}\quad \pmf{X_{\vi} |X_{\not \vi}} = \pmf{X_{\vi}|X_{\neigh{\vi}}},
 \end{equation}
 i.e., if the distribution of $X_{\vi}$ depends on the remaining RVs only via the RVs
 neighboring $\vi$.
 If $\pmf{X}$ is unspecified, but known to belong to a family of distributions for
 which~\eqref{eq:MRF} holds for every member, then we say that $X$ is a $\graph$-MRF.
\end{defin}

For any $A,B\subseteq \sets{V}$, the entropy of $X_{A}$ is defined as
\begin{align}
	\ent{X_{A}} := -\sum_{x_{A} \in \sets{X}_{A}} \pmf{X_{A}}(x_{A}) \log \pmf{X_{A}}(x_{A})
\end{align}
and the conditional entropy of $X_A$ given $X_B$ as
$\ent{X_{A} | X_{B}} :=\ent{X_{A\cup B}} - \ent{X_{B}}$. With this notation,
the lemma below follows immediately from Definition~\ref{def:MRF}.

\begin{lem}\label{lem:MIinMRF}
	$X$ is a $\graph$-MRF if and only if (iff), for every $\vi \in\sets{V}$,
	$\ent{X_{\vi}|X_{\not{\vi}}}=\ent{X_{\vi}|X_{\neigh{\vi}}}.$
\end{lem}

Note that if $X$ is a
$\graph$-MRF, then it is a MRF on every graph with vertices $\sets{V}$ whose edge set is a superset
of $E$. Trivially, every $X$ is a MRF on the complete graph. Of particular interest is thus the
\emph{minimal} graph w.r.t.\ which $X$ is a MRF. We will assume throughout this paper that the graph
$\graph$ w.r.t.\ which $X$ is a MRF is minimal. 

\section{Problem Statement and Motivating Examples}
\label{sec:formulation}

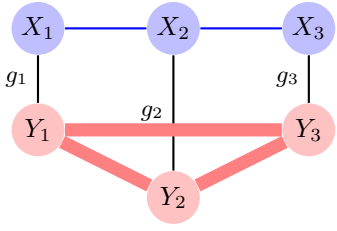
\begin{figure}
 \centering
\begin{tikzpicture}[scale=1.8, auto,swap]
    \foreach \pos/\name in {{(0,.75)/X_1}, {(1,.75)/X_2}, {(2,.75)/X_3}}
        \node[vertex] (\name) at \pos {$\name$};
    \foreach \pos/\name in {{(0,0)/Y_1}, {(1,-0.5)/Y_2}, {(2,0)/Y_3}}
        \node[selected vertex] (\name) at \pos {$\name$};
        
    \foreach \source/ \dest /\weight in {X_1/X_2/, X_2/X_3/}
        \path[edge, blue] (\source) -- node[weight] {$\weight$} (\dest);
    \foreach \source/ \dest /\weight in {X_1/Y_1/g_1,X_2/Y_2/g_2,X_3/Y_3/g_3}
        \path[edge] (\source) -- node[weight] {$\weight$} (\dest);
        
    \path [selected edge] (Y_1) -- (Y_2);
    \path [selected edge] (Y_2) -- (Y_3);
    \path [selected edge] (Y_1) -- (Y_3);
\end{tikzpicture}
\caption{The problem of lumpability. The blue vertices and edges correspond to the
	original $(\graph,\pmf{X})$-MRF $X$, the black labeled edges correspond to functions through which the
	RVs in the MRF are observed, thus defining $Y:=(g_1(X_1),g_2(X_2),g_3(X_3))$.
	In general, the
	minimal graph w.r.t.\ which $Y$ is a MRF is complete (see red vertices and edges): By observing a
	Markov path $X_1\text{---}X_2\text{---}X_3$ through a non-injective function, the Markov property is lost in general.
	The lumpability problem seeks conditions on $p_{X}$ and $\{g_{\vi}\}$ such that the minimal graph
	for $Y$ is equivalent to the original graph $\graph$ or a subgraph of $\graph$.
}
\label{fig:trianglegraph}
\end{figure}

In this work, we consider functions of MRFs. Specifically, let $\{g_{\vi}, i\in \sets{V}\}$
(subsequently abbreviated as $\{g_{\vi}\}$ to simplify notation) be a set of functions
$g_{\vi}{:}\ \sets{X}_{\vi}\to\sets{Y}_{\vi}$ indexed by the vertices $\vi \in\sets{V}$, and let
$Y_{\vi}:=g_{\vi}(X_{\vi})$.
For $A\subseteq \sets{V}$, we define the function $g_A{:}\ \sets{X}_A\to\sets{Y}_A$ as the functions
$g_{\vi}$, $\vi \in A$, applied to $X_A$ coordinate-wise, i.e.,
$g_{A}(X_{A}):=(g_{i}(X_{i}), i\in A) = Y_{A}$, and, as before, use the abbreviation
$g(X):=g_{\sets{V}}(X)=Y$.
We call a set of functions $\{g_{\vi}\}$ \emph{non-trivial} if at least one
function $g_{\vi}$ is non-injective. Given a $(\graph,\pmf{X})$-MRF $X$ and a set of functions
$\{g_{\vi}\}$, we call the tuple $(\graph,\pmf{X},\{g_{\vi}\})$ the \emph{lumping} of $X$. We will focus on
the following two problems:

\begin{problem}[Lumpability]
	\label{prob:lumpability}
	Determine conditions on the lumping $(\graph,\pmf{X},\{g_{\vi}\})$ so that $Y$ is a MRF w.r.t.
	$\graph$, where in this case we say $(\graph,\pmf{X},\{g_{\vi}\})$ is \emph{lumpable}, see
	Fig.~\ref{fig:trianglegraph}.
	By the remark below Lemma~\ref{lem:MIinMRF}, $(\graph,\pmf{X},\{g_{\vi}\})$ is lumpable
whenever it does not introduce new edges, i.e., 
whenever $Y$ is a $(\graph',\pmf{Y})$-MRF with $\graph'=(\sets{V},E')$ and $E'\subseteq E$
\end{problem}

\begin{problem}[Information Preservation]\label{prob:infopreservation}
	Determine conditions on the lumping $(\graph,\pmf{X},\{g_{\vi}\})$ so that $\ent{Y}=\ent{X}$,
	where in this case we say $(\graph,\pmf{X},\{g_{\vi}\})$ is \emph{information-preserving}.
\end{problem}

Throughout this work we assume 
the set of functions $\{g_{\vi}\}$ is non-trivial. Otherwise, if all the functions $g_{\vi}$
are injective, then $X$ and $Y$ would have the same distribution since $\{g_{i}\}$ is simply a relabeling of 
the distribution's domain, and so the lumping would be trivially lumpable and information preserving.
We also assume that $\graph$ is connected, which is
w.l.o.g. since the RVs of different components of the graph are independent, and this independence is retained
for any set of functions $\{g_{\vi}\}$. 

To get some intuition on why a function of a MRF may not be a MRF on the same graph, note that
$X_{\vi}$ and
$X_{\not \vi}$ are conditionally independent given $X_{\neigh{\vi}}$ only when $X_{\neigh{\vi}}$ contains
all the information about $X_{\vi}$ that is available in $X_{\not \vi}$. Taking a function of $X_{\neigh{\vi}}$ may reduce this information to a point where $Y_{\neigh{\vi}}$ no longer contains all the information
about $Y_{i}$ that is available in $Y_{\not \vi}$, which effectively introduces edges in the minimal graph for $Y$ that have 
not been present in $\graph$.
This parallels the fact that a function of a Markov chain rarely results in a Markov chain~\cite[Th.~31]{GurvitsLedoux_MarkovPropertyLinearAlgebraApproach}.
(A Markov chain is a $\graph$-MRF where $\graph$ is the infinite path graph, i.e., with the natural
numbers $\mathbb{N}$ as the set of vertices and $\{ \{i,i+1\} : i \in \mathbb{N} \}$ as the set of
edges.)
Regarding information-preservation, a lumping is information-preserving iff $\{g_{\vi}\}$
maps the support of $\pmf{X}$ injectively. Thus, both lumpability and information-preservation
appear to be the exception rather than the rule.
The following examples demonstrate different lumpability and information-preservation scenarios and
give some intuition on the corresponding lumpings $(\graph,\pmf{X},\{g_i\})$.

\begin{example}[Neither Information-Preserving nor Lumpable]\label{ex:nonIp_non_Lump}\label{ex:complete}
	Let $X_1\text{---}X_2\text{---}X_3$ be a Markov path, i.e.,  a  $\graph$-MRF on the path
	graph $\graph=(\{1,2,3\}, \{\{1,2\},\{2,3\}\})$,
	where each RV $X_{\vi}$ takes values from $\{0,1,2\}$. Suppose
 that $\pmf{X_1|X_2}(1|0)=\pmf{X_3|X_2}(1|2) = 0$, $\pmf{X_1|X_2}(1|2)=\pmf{X_3|X_2}(1|0) = p > 0$,
 and  $\pmf{X_2}(0)=\pmf{X_2}(2)\in(0,0.5)$.
For all other configurations, assume $p_{X_1|X_2}$ and $p_{X_3|X_2}$ are positive.
Let $g_{\vi}(x_{\vi})=\mathrm{mod}(x_{\vi},2)$ for every $\vi$,
then one can verify that
$\pmf{Y_1|Y_2}(1|0) = p/2 =  \pmf{Y_3|Y_2}(1|0)$, while 
$\pmf{Y_1,Y_3|Y_2}(1,1|0) = 0$. Thus, $Y_1$ and $Y_3$ are not conditionally independent given $Y_2$, and so the minimal graph for $Y$ contains the new edge $\{1,3\}$, i.e., the lumping
$(\graph,\pmf{X},\{g_{\vi}\})$ is not lumpable. (In this example the minimal graph for $Y$ is the
complete graph, see Fig.~\ref{fig:trianglegraph}.)
Furthermore, since, e.g., $x=(0,0,0)$ and $x'=(0,0,2)$ both have
positive probabilities, but are mapped to the same $y=(0,0,0)$, the lumping is not information-preserving.
\end{example}

\begin{example}[Information-Preserving but not Lumpable]\label{ex:ip_nonLump}
Let $X_1:=X_2+Z_1$ and $X_3:=X_2+Z_3$, where $Z_1\in\{0,1\}$, $X_2\in\{-1,1\}$, and $Z_3\in
\{-1,0\}$ are mutually independent RVs.  It follows that $X_1\text{---}X_2\text{---}X_3$ is a Markov
path as in the previous example with edges $E = \{\{1,2\}, \{2,3\}\}$.
Assume $g_1$ and $g_3$ are the
identity functions and $g_2\equiv 0$.  Since $Y_2$ is constant, $Y_1$ and $Y_3$ are conditionally
independent given $Y_2$ iff $Y_1$ and $Y_3$ are independent, which is not true due to the coupling
through $X_2$. (Assuming $p_{X_2}$ is strictly positive.) Hence, the lumping
$(\graph,\pmf{X},\{g_{\vi}\})$ is not lumpable since the minimal graph for $Y$ must contain the edge
$\{1,3\}$, which is not in $E$. (Indeed, $Y$ is a MRF w.r.t.\ the graph $(\{1,2,3\},\{\{1,3\}\})$.)
Furthermore, one can show that $X_2=1$ iff $X_1>0$ and that $X_2=-1$ iff
$X_3<0$, hence $Y=(X_1,0,X_3)$ contains the same information as $X$, i.e., the lumping is
information-preserving.
\end{example}

\begin{example}[Lumpable and Information-Preserving]\label{ex:ip_lump}
Let $X_2:=(X_1,Z_2,X_3)$, where  $X_1$, $Z_2$, and $X_3$ are mutually independent.
Then, we have the Markov path $X_1\text{---}X_2\text{---}X_3$ again,
where the PMF $\pmf{X}$ satisfies
\begin{multline}
    \pmf{X}(x_1,(z_1,z_2,z_3),x_3) \\= \begin{cases}
     \pmf{X_1}(x_1)\pmf{Z_2}(z_2)\pmf{X_3}(x_3), & x_1=z_1, x_3=z_3\\
     0, &\text{else}.
    \end{cases}
\end{multline}
Now suppose that $g_1$ and $g_3$ are the identity mappings and that $g_2$ is such that
$g_2(z_1,z_2,z_3)=z_2$. Obviously, the thus defined RVs $Y_1$, $Y_2$, and $Y_3$ are independent,
i.e., $Y$ is a MRF on the empty graph, and so $(\graph, p_{X}, \{g_{\vi}\})$ is lumpable.
Furthermore, it is clear that $\ent{g(X)}=\ent{X}$, and so the lumping is information-preserving.
\end{example}

\section{Previous Work on MRFs}
\label{sec:related}

Yeung et al.\ characterized MRFs using the $I$-measure~\cite{Yeung_ISIT,Yeung_IT}. Specifically,
if $X$ is a $\graph$-MRF and $A\subseteq \sets{V}$, they investigated the minimal graph
$\graph_A=(A,E_A)$ on which $X_A$ is a MRF. They showed that $E_A$ contains $\{\vi,\vj\}\in[A]^2$ if
either $\{\vi,\vj\}\in E$ or if there is a path between $\vi$ and $\vj$ in $\graph$ of which all intermediate
vertices lie in $\sets{V}\setminus A$, see \cite[Th.~5]{Yeung_ISIT} or \cite[Th.~8]{Yeung_IT}.
More generally, Sadeghi~\cite{Sadeghi_Marginalization}
characterized probabilistic graphical models, admitting mixed graphs $\tilde\graph$ with directed,
doubly-directed, and undirected edges, and presented an algorithm that generates a corresponding
graph for a subset $A\subseteq \sets{V}$ of the vertices of
$\tilde\graph$,~cf.~\cite[Algorithm~1]{Sadeghi_Marginalization}. With the restriction to undirected
graphs, this algorithm terminates with $\graph_A$ as discussed in~\cite{Yeung_IT}.
Much earlier, P\'erez and Heitz investigated this problem from a Gibbs field perspective, i.e.,
using potential functions, where they showed that $X_A$ is a
$(\graph_A,\pmf{X_A})$-MRF~\cite[Th.~2]{Perez_TIT}, but that $\graph_A$ is only minimal if
additional conditions are fulfilled~\cite[Th.~3]{Perez_TIT}.

Below we clarify some connections between previous works and the current one.
Given a MRF $X$ w.r.t. a graph $\graph_{X}$ and a transformation $p_{Y|X}$ of $X$ to $Y$,
assume the joint RV $(X,Y)$ is a MRF on a graph $\graph_{X,Y}$. (According to the problem
formulation in Problem~\ref{prob:lumpability}, such a graph is not needed
in the current paper and only assumed in this paragraph to facilitate discussions relative
to previous works.)
The vertex set of this graph is the disjoint union of the vertices of $\graph_{X}$ and a set of vertices
associated with $Y$, and the edge set is obtained from the edges of $\graph_{X}$ and the transformation $p_{Y|X}$.
Determining on which graph $\graph_Y$ the RV $Y$ is a MRF can then be done by
applying~\cite[Th.~5]{Yeung_ISIT} or \cite[Th.~2~\&~3]{Perez_TIT} to $\graph_{X,Y}$ for the subset of vertices that are
associated with $Y$. With this setup, the primary distinctions between previous works and the
current one are the following:
\cite{Yeung_ISIT, Yeung_IT} make no assumptions on $p_{Y|X}$, \cite{Perez_TIT} assumes $p_{Y|X}$ is
strictly positive, and this work assumes
\begin{equation}
	\label{eq:cond-trans}
	p_{Y|X}(y|x) = \prod_{\vi\in\sets{V}} \mathbb{I}[g_{\vi}(x_{\vi})=y_{\vi}],
\end{equation}
where $\mathbb{I}[\cdot]$ is the indicator function, i.e., $p_{Y|X}$ factors as the product of
degenerate distributions $p_{Y_{\vi}|X_{\vi}}$ that account to the fact that $Y_{\vi}$ is a deterministic function of $X_{\vi}$.

Unfortunately, Problem~\ref{prob:lumpability} cannot be solved with the framework
in~\cite{Perez_TIT} since the conditional distribution \eqref{eq:cond-trans} is not strictly positive, nor can
it be solved using~\cite{Yeung_ISIT,Yeung_IT} since the framework therein finds a graph 
$\graph_{Y}$ that is minimal for any $\{g_{\vi}\}$ (in fact for any $p_{Y_{\vi}|X_{\vi}}$) and any
$p_{X}$ in the family of distributions specified by $\graph_{X}$. In contrast, here we are given a
fixed set of transformations $\{g_{\vi}\}$ and (often) a fixed distribution $p_{X}$.
Indeed, if $\graph_{X}$ is connected, then~\cite[Th.~5]{Yeung_ISIT} leads to $\graph_Y$ being complete. 
In other words, for any MRF $X$ on a connected graph, \cite[Th.~5]{Yeung_ISIT} states that one
can find a PMF $\pmf{X}$ and a set of functions $\{g_{\vi}\}$ 
(more precisely $p_{Y_{\vi}|X_{\vi}}, i = 1, \dots, N,$ as the theorem does not assume deterministic mappings)
such that $Y$ does not satisfy any
conditional independence statements.
(See Example~\ref{ex:complete} for an explicit choice of $p_{X}$ and $\{g_{\vi}\}$ that results in
the complete graph in the case of
the Markov path.)

Little work has been done regarding information-preserving lumpings of a MRF, see
Problem~\ref{prob:infopreservation}.
A work in a related direction is~\cite{Reyes_EntropyBounds}, which shows that under certain conditions the entropy
$\ent{X_A}$, for $A \subseteq \sets{V}$, can be bounded from above by the entropy of a MRF w.r.t.\ the subgraph of $\graph$
induced by $A$.

\section{Sufficient Conditions for MRF Lumpability}\label{sec:sufficient}
Below we investigate  Problem~\ref{prob:lumpability}, namely, we determine sufficient conditions for
the lumping $(\graph,\pmf{X},\{g_{\vi}\})$ to be lumpable.
Note that according to Problem~\ref{prob:lumpability}, $(\graph,\pmf{X},\{g_{\vi}\})$ is
lumpable if $Y$ is a $(\graph,\pmf{Y})$-MRF, even if $\graph$ is not minimal for $Y$. 
We further assume within this section that $\pmf{X}(x)>0$ for every $x\in\sets{X}$, and that $g$ is surjective, i.e., $\mathcal{Y}$ is the image of $\mathcal{X}$ under $g$. This allows the characterization of
a MRF via its connection to Gibbs fields. (Despite this assumption, the joint distribution $p_{XY}$ is not strictly
positive, see \eqref{eq:cond-trans}.)  Specifically, let $\psi_A{:}\ \sets{X}_A\to\reals$ be a potential function. We abuse
notation and extend the domain of $\psi_A$ to $\sets{X}$, i.e., for $x=(x_1,\dots,x_N)\in\sets{X}$
we write $\psi_A(x):=\psi_A(x_A)$, where $x_A:=(x_{\vi},\ \vi \in A)$.
The following lemma gives the characterization required in this section.
\begin{lem}[Hammersley-Clifford~\cite{hammersley-clifford}]\label{lem:HC} $X$ is a $(\graph,\pmf{X})$-MRF satisfying
	$\pmf{X}(x)>0$ for every $x\in\sets{X}$ iff there exists a family of potential functions
	$\{\psi_C,\ C\in\sets{C}\}$ such that
  \begin{subequations}
    \begin{equation}
      \forall x\in\sets{X}{:}\quad \pmf{X}(x) = \frac{1}{Z} \prod_{C\in\sets{C}} \psi_C(x),
		\label{eq:hc}
  \end{equation}
  where
  $\sets{C}$ is the set of cliques of $\sets{G}$ and
  \begin{equation}\label{eq:partitionfunction}
	  Z := \sum_{x\in\sets{X}} \prod_{C\in\sets{C}} \psi_C(x).
  \end{equation}
  \end{subequations}
\end{lem}
 
Since the potential functions in the family $\{\psi_C,\ C\in\sets{C}\}$ are defined on cliques, we
call $\psi_C$ a \emph{clique potential}. Note that the choice of $\{\psi_C,\ C\in\sets{C}\}$ is not
unique. Indeed, Lemma~\ref{lem:HC} may be satisfied with a subset of potential functions being
identically one.

For a non-trivial set of functions $\{g_{\vi}\}$, $Y$ is a $(\graph,\pmf{Y})$-MRF iff we can find a family of potential functions $\{U_C,\ C\in\sets{C}\}$ such that, for every $y\in\sets{Y}$
\begin{multline}\label{eq:cliqueY}
    Z\cdot\pmf{Y}(y) = Z\cdot\sum_{x\in g^{-1}(y)} \pmf{X}(x) \\=\sum_{x\in g^{-1}(y)} \prod_{C\in\sets{C}} \psi_C(x)
    = \prod_{C\in\sets{C}} U_C(y)
\end{multline}
where $Z$ is the partition function from~\eqref{eq:partitionfunction}.
Such a family can obviously be found if, for all $y \in \sets{Y}$, the family
$\{\psi_C,\ C\in\sets{C}\}$ is constant on the preimage $g^{-1}(y):=\{x \in \sets{X} : g(x) = y\}$.
Specifically, if for every $C\in\sets{C}$ and for every $y\in\sets{Y}$ we have
\begin{equation}\label{eq:allequal}
    \psi_C(x) = \psi_C(x'), \ \forall x,x'\in g^{-1}(y),
\end{equation}
then we can choose $U_C(y)$ as this common value multiplied by the cardinality of the preimage
	$g^{-1}(y)$ to satisfy~\eqref{eq:cliqueY}. 
The remainder of this section will give milder conditions than \eqref{eq:allequal} that guarantee lumpability. 

For any clique $C$ that contains vertex $i$, we say $\psi_{C}$  depends on $x_{\vi}$ \emph{only via}
$y_{i}$ if for all $y_{\vi}\in\sets{Y}_{\vi}$ and $x_{\vi},x_{\vi}'\in g_{\vi}^{-1}(y_{\vi})$ 
 \begin{align}
	 \label{eq:only-via}
	 \psi_C(x_{\not \vi},x_{\vi}) = \psi_C(x_{\not \vi}, x'_{\vi}),
     \forall x_{\not \vi}\in\sets{X}_{\not \vi},
 \end{align}
otherwise, we say $\psi_C$ \emph{strictly} depends on $x_{\vi}$.
The following result will assume that for every vertex $\vi$ there is at most one clique potential
that is allowed to strictly depend on $x_{\vi}$. 
    For all $\vi$, let $C'(i)$ denote the corresponding clique.
	(If no potential function strictly depends on $x_{\vi}$ then $C'(i)$ is chosen as
	any clique involving $\vi$.)
	We can view this as a mapping $C'{:}\ \sets{V}\to\sets{C}$ that assigns to each vertex $i$ the unique
	clique that may strictly depend on $x_{i}$, which in effect partitions $\sets{V}$
	into equivalence classes $\sets{V}_{\cl}$, $\cl = 1, \dots, \cL$, such that all the vertices
	$i \in \sets{V}_{\cl}$ are assigned the same clique $C'(i)$. For convenience, the clique
	$C'(i)$, common to all $i\in \sets{V}_{\cl}$, will be denoted $C'(\sets{V}_{\cl})$.

\begin{prop}\label{prop:Gibbs}
 Assume $X$ is a $(\graph,\pmf{X})$-MRF characterized by a family $\{\psi_C,\ C\in\sets{C}\}$ of
 potential functions such that, for all $i \in \sets{V}$, there is at most one clique whose potential may strictly depend on $x_{\vi}$, then $Y$ is a $(\graph,\pmf{Y})$-MRF.

 Moreover, with $C'$ and $\sets{V}_{1}, \dots, \sets{V}_{L}$ as above,
 the $(\graph,\pmf{Y})$-MRF is characterized by the family $\{U_C,\ C\in\sets{C}\}$ of potential functions, where
\begin{subequations}
\begin{align}
	& U_{C'(\sets{V}_{\cl})}(g(x)) = \kern -.5em \sum_{x_{\sets{V}_{\cl}}'\in
	g_{\sets{V}_{\cl}}^{-1}(g_{\sets{V}_{\cl}}(x_{\sets{V}_{\cl}}))} \kern -.5em
	\psi_{C'(\sets{V}_{\cl})}(x_{\sets{V}_{\cl}}',x_{\sets{V}\setminus\sets{V}_{\cl}})\label{eq:lumpability:1st} 
\end{align}
for $\cl=1,\dots,\cL$, and
\begin{align}
	U_C(g(x)) = \psi_C(x), 
	 \forall C\in\sets{C}\setminus\cup_{\vj\in\sets{V}} C'(\vj).
\end{align}
\end{subequations}
\end{prop}

\begin{cor}
  If~\eqref{eq:allequal} holds, then Proposition~\ref{prop:Gibbs} is trivially fulfilled. In this
  case, $C'(\vi)$ is any clique of which $\vi$ is a member and~\eqref{eq:lumpability:1st} simplifies to
\begin{equation}
	U_{C'(\sets{V}_{\cl})}(g(x)) = |g_{\sets{V}_{\cl}}^{-1}(g_{\sets{V}_{\cl}}(x_{\sets{V}_{\cl}}))|\cdot\psi_{C'(\sets{V}_{\cl})}(x).
\end{equation}
\end{cor}

Since, even for a fixed joint PMF $\pmf{X}$, the family of potential functions is not unique, $Y$ is
a $(\graph,\pmf{Y})$-MRF if we can find at least one family of potential functions that characterizes
$\pmf{X}$ and for which Proposition~\ref{prop:Gibbs} holds.

\begin{example}\label{ex:nonunique}
	Let $X_1\text{---}X_2\text{---}X_3$ be a Markov path and fix a set of functions $\{g_1, g_2, g_3\}$.  Suppose that
	$\psi_{\{\vi\}}$, for $\vi=1,2,3$, are arbitrary, and
	$\psi_{\{1,2\}}(x_1,x_2)=U_{\{1,2\}}(g_1(x_1),g_2(x_2))$ and
	$\psi_{\{2,3\}}(x_2,x_3)=U_{\{2,3\}}(g_2(x_2),g_3(x_3))$
	for some $U_{\{1,2\}}$ and $U_{\{2,3\}}$.
	Thus, only $\psi_{\{\vi\}}$ may strictly depend on $x_{\vi}$, and so
	Proposition~\ref{prop:Gibbs} applies.
	Now,  the same PMF $\pmf{X}$ can be characterized using the potentials
	$\psi'_{\{1,2\}}=\psi_{\{1,2\}}\cdot\sqrt{\psi_{\{2\}}}$,
	$\psi'_{\{2,3\}}=\psi_{\{2,3\}}\cdot\sqrt{\psi_{\{2\}}}$, 
	$\psi'_{\{1\}}=\psi_{\{1\}}$,
	$\psi'_{\{2\}}=1$, and
	$\psi'_{\{3\}}=\psi_{\{3\}}$.
	Assuming $\psi_{2}$ strictly depends on $x_{2}$, then both $\psi'_{\{1,2\}}$ and $\psi'_{\{2,3\}}$ strictly depend on
	$x_2$, and so the condition in Proposition~\ref{prop:Gibbs} is violated.
\end{example}

We now complement Proposition~\ref{prop:Gibbs}, which is based on clique potentials, by a sufficient condition for lumpability based on conditional entropies. This condition follows from Lemma~\ref{lem:MIinMRF}, the data processing inequality, and
the fact that conditioning reduces entropy.
\begin{prop}\label{prop:IT}
	Let $X$ be a $\graph$-MRF. 
	If, for every $\vi \in\sets{V}$,
 \begin{equation}\label{eq:ITsuff}
	 \ent{Y_{\vi}|Y_{\neigh{\vi}}} = \ent{Y_{\vi}|X_{\neigh{\vi}}}
 \end{equation}
 then $Y$ is a $\graph$-MRF.
\end{prop}

Equation~\eqref{eq:ITsuff} gives an intuitive interpretation for lumpability of MRFs: If (but not
only if, see Example~\ref{ex:ITcounter} below) the neighbors of $X_{\vi}$ are not more informative about
the outcome of $Y_{\vi}$ than the function of these neighbors, then $Y$ is a $\graph$-MRF. In other
words, $Y$ is a $\graph$-MRF if the lumping is such that $Y_{\neigh{\vi}}$ captures all information in
$X_{\neigh{\vi}}$ that is relevant to $Y_{\vi}$.

\begin{example}\label{ex:ITcounter}
 Let $X=(X_1,X_2)$ be a Markov path, i.e., $\sets{V}=\{1,2\}$ and $E=\{1,2\}$. Trivially, since
 $\graph$ is the complete graph, $Y$ is a $\graph$-MRF for every set of functions $\{g_1,g_2\}$.
 However, one can construct examples for $\pmf{X}$ and $\{g_1,g_2\}$ such that there exists
 $y\in\sets{Y}$ and a pair $x_1,x_1'\in g_1^{-1}(y_1)$ such that
 \begin{equation}
     \pmf{Y_2|X_1}(y_2|x_1) \neq \pmf{Y_2|X_1}(y_2|x_1').
 \end{equation}
 Thus, the condition of Proposition~\ref{prop:IT} does not hold, showing that it is only sufficient but not necessary.
\end{example}

There is some similarity between~\eqref{eq:ITsuff} and an information-theoretic
sufficient condition for the lumpability of an irreducible and aperiodic Markov chain
$X_1$---$X_2$---$X_3$---$\cdots$ (see~\cite{Kemeny_FMC} for terminology). 
Suppose that $X$ is stationary, i.e., the alphabets of $X_{\vi}$ are all the same,
$\pmf{X_{\vi+1}|X_{\vi}}=\pmf{X_{\vj+1}|X_{\vj}}$ for every $\vi,\vj\in\mathbb{N}$, and the initial distribution
$\pmf{X_1}$ coincides with the unique distribution invariant under the one-step conditional
distribution $\pmf{X_{\vi+1}|X_{\vi}}$. If further all the functions $g_{\vi}$ are identical, i.e.,
$g_{\vi}=g_{0}$, $\vi\in\sets{V}$, for some function $g_{0}$, then one can show
that the tuple $(\graph,\pmf{X},g_0)$ is lumpable if~\cite[Th.~2]{GeigerTemmel_kLump}
\begin{equation}\label{eq:ITEq:Lump}
	\ent{Y_{\vi}|X_{\vi-1}} = \ent{Y_{\vi}|Y_{\vi-1}}.
\end{equation}
(By stationarity, it suffices that~\eqref{eq:ITEq:Lump} holds for any $\vi$.)
The main difference between~\eqref{eq:ITsuff} and~\eqref{eq:ITEq:Lump} is that the
latter is conditioned on only a subset of the neighbors, which corresponds to the case in which 
$\graph$ is directed, i.e., for $X_1\rightarrow X_2 \rightarrow \cdots$. Proposition~\ref{prop:IT}
shows that, for undirected graphs,~\eqref{eq:ITsuff} takes the place
of~\eqref{eq:ITEq:Lump} in a sufficient condition for lumpability.

\iflong
{We end this section with a sufficient condition for \eqref{eq:ITsuff} to hold for a MRF with a
strictly positive $p_{X}$.
Namely, while Proposition~\ref{prop:Gibbs} restricts the number of cliques whose potential functions strictly
depend on $x_{\vi}$; it does not restrict the number of components of $x$ on which the potential
function of a given clique may strictly depend on. The following proposition 
limits the number of components (of $x$) a clique potential may strictly depend on to one.

\begin{prop}\label{prop:ITGibbs}
	Let $X$ be a $(\graph,\pmf{X})$-MRF as in Proposition~\ref{prop:Gibbs} with 
	$C'(\vi)\neq C'(\vj)$ for every pair of
distinct vertices $\vi,\vj\in\sets{V}$,
then
\begin{equation}\label{eq:entropies_equal}
	\ent{Y_{\vi}|Y_{\neigh{\vi}}} = \ent{Y_{\vi}|X_{\neigh{\vi}}}, \ \forall i \in \sets{V}.
\end{equation}
\end{prop}
}
\fi

\section{Information-Preserving MRF Lumpings}\label{sec:loss}
We next briefly talk about information-preserving lumpings of MRFs,
see~Problem~\ref{prob:infopreservation}. A lumping can only be information-preserving if $g$ maps
the support of $\pmf{X}$ injectively. If the support of $\pmf{X}$ coincides with $\sets{X}$, then
only trivial sets of functions $\{g_{\vi}\}$, in which every $g_{\vi}$ is injective, can be
information-preserving. In this section, we therefore drop the assumption that $\pmf{X}$ is positive
on $\sets{X}$. However, while it is clear that $\ent{X}=\ent{Y}$ iff $g$ is injective on the
support of $\pmf{X}$, this does not imply that every $g_{\vi}$ is injective on the support of
$\pmf{X_{\vi}}$.
In other words, a lumping $(\graph,\pmf{X},\{g_{\vi}\})$ can be information-preserving even if some or
all of the functions $g_{\vi}$ are non-injective, i.e., even if $\ent{X_{\vi}}>\ent{Y_{\vi}}$ for some
$\vi \in\sets{V}$.

\begin{prop}\label{prop:infopreservation}
	Let $X$ be a $(\graph,\pmf{X})$-MRF.
	\begin{subequations}
	\begin{itemize}
		\item For any graph $\graph$, if the lumping $(\graph,\pmf{X},\{g_{\vi}\})$ is information-preserving, then 
		\begin{equation}\label{eq:it:necessary}
			\forall \vi \in\sets{V}{:}\quad \ent{X_{\vi}|Y_{\vi},X_{\neigh{\vi}}}=0.
		\end{equation}
	 
	\item For any chordal graph $\graph$, the lumping $(\graph,\pmf{X},\{g_{\vi}\})$ is
		 information-preserving if there exist a vertex permutation $\perm_1,\dots,\perm_N$  and sets
		 $A_{\perm_i}=\neigh{\perm_i}\cap \{\perm_1,\dots,\perm_{i-1}\}$ such that
		 \begin{equation}\label{eq:it:sufficient}
			 \forall i\in\sets{V}{:}\quad \ent{X_{\perm_i}|Y_{\perm_i},X_{A_{\perm_i}}}=0.
		 \end{equation}
	\end{itemize}
	\end{subequations}
\end{prop}
 
\begin{example}\label{ex:infoloss}
 Let $X_1=X_2$, i.e., $X$ is a MRF on a path, which is a chordal graph. Assume that $g_1\equiv g_2$
 and that $g=(g_1,g_2)$ is non-injective on the support of $\pmf{X}$. Thus, $\ent{g(X)} < \ent{X}$.
 
 Since $\ent{X_1|X_2}=0$ and $\ent{X_2|X_1}=0$, we have $\ent{X_1|g_1(X_1),X_2}=0$ and
 $\ent{X_2|g_2(X_2),X_1}=0$, i.e., the necessary condition for information
 preservation~\eqref{eq:it:necessary} holds. However, we have that $\ent{X_1|Y_1}>0$ due to the
 non-injectivity of $g_1$, and so~\eqref{eq:it:sufficient} does not hold for the permutation
 $(v_{1}, v_{2}) = (1,2)$. A similar argument holds for the permutation $(v_{1}, v_{2}) = (2,1)$.
 Thus, the sufficient condition for chordal
 graphs~\eqref{eq:it:sufficient} is violated.
\end{example}

\begin{remark}
Let $X_1$---$X_2$---$X_3$---$\cdots$ be an irreducible, aperiodic, and stationary Markov chain. The
graph w.r.t.\ which $X$ is a MRF is an (infinite) path, which is chordal. Since $A_i=\{i-1\}$
(under the choice $v_{i} = i$ for all $i$) and due to
stationarity, the sufficient condition in~\eqref{eq:it:sufficient} simplifies to
$\ent{X_2|Y_2,X_1}=0$. We thus recover~\cite[Prop.~4]{GeigerTemmel_kLump}.
\end{remark}

While the condition that $g$ maps the support of $\pmf{X}$ injectively is an equivalent
characterization of information-preservation, the conditions in
Proposition~\ref{prop:infopreservation} (that are only necessary or sufficient) have practical
justification. Indeed, for alphabets $\sets{X}_{\vi}$ with fixed cardinality, the support of
$\pmf{X}$ grows exponentially in the number $N$ of vertices. In contrast,~\eqref{eq:it:necessary}
requires checking whether $g_{\vi}$ maps the support of $\pmf{X_{\vi}|X_{\neigh{\vi}}}$ injectively
for every $\vi$; the number of parameters characterizing this conditional PMF is exponential only in
the size of the neighborhood of $\vi$, which is much smaller than $N$ for sparse graphs.
Thus, rather than checking $g$ \emph{globally}, which is exponential in $N$, it
suffices to check a computationally less expensive \emph{local} condition for each $g_{\vi}$.

We finally remark that Proposition~\ref{prop:infopreservation} holds regardless whether $Y$ is a $\graph$-MRF or not, i.e., whether $(\graph,\pmf{X},\{g_{\vi}\})$ is lumpable or not. A better
understanding of the interactions between lumpability and information-preservation, i.e., between Problems~\ref{prob:lumpability} and~\ref{prob:infopreservation}, seems to be of practical and theoretical interest. Thus, a closer investigation of these interactions shall be the subject of future work.

\iflong
\section{Proofs}
\label{sec:proofs}

\subsection{Proof of Proposition~\ref{prop:Gibbs}}\label{proof:Gibbs}
First, note that if for $C\in\sets{C}$ and $\vi,\vj\in\sets{V}$ the clique potential $\psi_C$ is
constant on the preimages under $g_{\vi}$ and $g_{\vj}$, then $\psi_C$ is also constant on the
Cartesian product of these preimages.
%Indeed, if for all
%$x_{\sets{V}\setminus\{\vi,\vj\}}\in\sets{X}_{\sets{V}\setminus\{\vi,\vj\}}$, for all
%$y_{\vi}\in\sets{Y}_{\vi}$,
%$y_{\vj}\in\sets{Y}_{\vj}$, and all $x_{\vi},x_{\vi}'\in g_{\vi}^{-1}(y_{\vi})$,
%$x_{\vj},x_{\vj}'\in g_{\vj}^{-1}(y_{\vj})$ 
%we have 
% \begin{subequations}\label{eq:projections}
% \begin{IEEEeqnarray}{rcl}
%	 \psi_C(x_{\vi},x_{\vj},x_{\sets{V}\setminus\{\vi,\vj\}}) &=& \psi_C(x_{\vi}',x_{\vj},x_{\sets{V}\setminus\{\vi,\vj\}})\\
%	 \psi_C(x_{\vi},x_{\vj},x_{\sets{V}\setminus\{\vi,\vj\}}) &=& \psi_C(x_{\vi},x_{\vj}',x_{\sets{V}\setminus\{\vi,\vj\}}),
% \end{IEEEeqnarray}
% then we also have
% \begin{multline}
%	 \psi_C(x_{\vi},x_{\vj},x_{\sets{V}\setminus\{\vi,\vj\}}) = \psi_C(x_{\vi}',x_{\vj},x_{\sets{V}\setminus\{\vi,\vj\}})\\
%	 = \psi_C(x_{\vi},x_{\vj}',x_{\sets{V}\setminus\{\vi,\vj\}}) = \psi_C(x_{\vi}',x_{\vj}',x_{\sets{V}\setminus\{\vi,\vj\}}).
% \end{multline}
% \end{subequations}
Indeed, if this is the case, then for all
$x_{\sets{V}\setminus\{\vi,\vj\}}\in\sets{X}_{\sets{V}\setminus\{\vi,\vj\}}$, for all
$y_{\vi}\in\sets{Y}_{\vi}$,
$y_{\vj}\in\sets{Y}_{\vj}$, and all $x_{\vi},x_{\vi}'\in g_{\vi}^{-1}(y_{\vi})$,
$x_{\vj},x_{\vj}'\in g_{\vj}^{-1}(y_{\vj})$ 
we have 
 \begin{align}\label{eq:projections}
	 \psi_C(x_{\sets{V}}) = \psi_C(x_{\vi}',x_{\sets{V}\setminus\{\vi\}})
	 %= \psi_C(x_{\vi},x_{\vj}',x_{\sets{V}\setminus\{\vi,\vj\}}) =
	 = \psi_C(x_{\vi}',x_{\vj}',x_{\sets{V}\setminus\{\vi,\vj\}}),
 \end{align}
 where the first and second equalitys follow from the assumption that $\psi_{C}$ is constant on the
 preimages under $g_{i}$ and $g_{j}$, respectively. Thus, if $\psi_{C}$ depends on $x_i$ only via $y_i$ and on $x_j$ only via $y_j$, then it also depends on $x_{\{i,j\}}$ only via $y_{\{i,j\}}$.
 
We write
\begin{multline}
Z\cdot \pmf{Y}(y) \\
= \sum_{x\in g^{-1}(y)} \prod_{C'(\vi){:} \vi\in\sets{V}} \psi_{C'(\vi)}(x)
\prod_{C\in\sets{C}\setminus \cup_{\vj\in\sets{V}}C'(\vj)} \psi_C(x)
\end{multline}
%where the second product is a product over cliques the potentials of which are constant on the preimages under $g$. Furthermore, for the second product we note that, since the clique potential $\psi_C$ is constant on the preimages of $y$ under $g$, we can define a potential function $U_C{:}\ \sets{Y}\to\reals$ via setting $U_C(g(x)):=\psi_C(x)$. Thus, we get
where the second product is a product over cliques whose potentials are constant on the preimages
under $g$. In other words, each potential in the second product depends on $x$ only via $y$, and so
we can define a potential function $U_C{:}\ \sets{Y}\to\reals$ by setting $U_C(g(x)):=\psi_C(x)$.
From this, we have 
\begin{IEEEeqnarray*}{rcl}
    Z\cdot \pmf{Y}(y) %&& \notag\\
  \ &=& \kern-.5em\prod_{C\in\sets{C}\setminus \cup_{\vj\in\sets{V}}C'(\vj)} U_C(y)
  \times \sum_{x\in g^{-1}(y)} \prod_{C'(\vi){:} \vi\in\sets{V}} \psi_{C'(\vi)}(x) \notag\\
	&=& \kern-.5em\prod_{C\in\sets{C}\setminus \cup_{\vj\in\sets{V}}C'(\vj)} U_C(y)
	\kern.2em \times
	\notag \\ &&
	\sum_{x_{\sets{V}_1}\in g_{\sets{V}_1}^{-1}(y_{\sets{V}_1})} \cdots
	\sum_{x_{\sets{V}_\cL}\in g_{\sets{V}_\cL}^{-1}(y_{\sets{V}_\ell})} \prod_{C'(\vi){:}
	\vi\in\sets{V}} \psi_{C'(\vi)}(x)\\
	&=& \kern-.5em \prod_{C\in\sets{C}\setminus \cup_{\vj\in\sets{V}}C'(\vj)} U_C(y)
	\kern.2em \times
	\notag \\ &&
	\kern1.8em
	\prod_{\cl=1}^\cL \sum_{x_{\sets{V}_{\cl}}\in
	g_{\sets{V}_{\cl}}^{-1}(y_{\sets{V}_{\cl}})}
	\psi_{C'(\sets{V}_{\cl})}(x_{\sets{V}_{\cl}},x_{\sets{V}\setminus\sets{V}_{\cl}}'),
\end{IEEEeqnarray*}
where $x_{\sets{V}\setminus\sets{V}_{\cl}}'$ is such that
$g_{\sets{V}\setminus\sets{V}_{\cl}}(x_{\sets{V}\setminus\sets{V}_{\cl}}')=y_{\sets{V}\setminus\sets{V}_{\cl}}$,
and 
the last equality follows from~\eqref{eq:projections} since $\psi_{C'(\sets{V}_{\cl})}$
%depends on $x_{\sets{V}\setminus\sets{V}_{\cl}}$ only via $y_{\sets{V}\setminus\sets{V}_{\cl}}$
depends on $x_{\vi}$ only via $y_{\vi}$ for all $\vi \in \sets{V}\setminus\sets{V}_{\cl}$.
%
%This allows us to define clique potentials
Define clique potentials
$U_{C'(\sets{V}_{\cl})}{:}\ \sets{Y}\to\reals$ by setting
\begin{equation}\label{eq:proof1:UC}
	U_{C'(\sets{V}_{\cl})}(g(x)) := \sum_{x_{\sets{V}_{\cl}}'\in
	g_{\sets{V}_{\cl}}^{-1}(g_{\sets{V}_{\cl}}(x_{\sets{V}_{\cl}}))}
	\psi_{C'(\sets{V}_{\cl})}(x_{\sets{V}_{\cl}}',x_{\sets{V}\setminus\sets{V}_{\cl}}),
\end{equation}
then we have
\begin{align*}
  Z\cdot \pmf{Y}(y)
  = \kern-1.2em \prod_{C\in\sets{C}\setminus \cup_{\vi\in\sets{V}}C'(\vi)} \kern-1.4em U_C(y) \times \prod_{\cl=1}^L U_{C'(\sets{V}_{\cl})}(y) 
  = \prod_{C\in\sets{C}} U_C(y),
\end{align*}
which completes the proof.\qed

\subsection{Proof of Proposition~\ref{prop:IT}}\label{proof:IT}
Since $X$ is a $\graph$-MRF and $Y_{\vi}$ is a function of $X_{\vi}$, we have
$\ent{Y_{\vi}|X_{\neigh{\vi}}}=\ent{Y_{\vi}|X_{\not{\vi}}}\le\ent{Y_{\vi}|Y_{\not{\vi}}}\le\ent{Y_{\vi}|Y_{\neigh{\vi}}}$,
where the first and second inequalities are due to the data processing inequality and the fact that
conditioning reduces the entropy,
respectively. Thus, given \eqref{eq:ITsuff}, the above inequalitis hold with equality, i.e., we have
$\ent{Y_{\vi}|Y_{\not{\vi}}}=\ent{Y_{\vi}|Y_{\neigh{\vi}}}$.
Lemma~\ref{lem:MIinMRF} completes the proof.
\qed

\subsection{Proof of Proposition~\ref{prop:ITGibbs}}\label{proof:ITGibbs}
 We have, for all $x\in\sets{X}$,
 \begin{align}
	 \label{eq:pX-Ni}
	 \pmf{X_{\vi}|X_{\not \vi}}(x_{\vi}|x_{\not \vi}) 
	 & =
	 \frac{\prod_{C\in\sets{A}}\psi_C(x)}{\sum_{x_{\vi}'\in\sets{X}_{\vi}}\prod_{C\in\sets{A}}\psi_C(x_{\vi}',x_{\not \vi})}
	 \nonumber \\
	 & =
	 \pmf{X_{\vi}|X_{\neigh{\vi}}}(x_{\vi}|x_{\neigh{\vi}}),
 \end{align}
 where $\sets{A}\subseteq\sets{C}$ is the set of cliques containing vertex $\vi$. The first equality
 is by \eqref{eq:hc} and the definition of the conditional distribution, and the second equality is by
 the definition of a MRF \eqref{eq:MRF}.
 Fix $y\in\sets{Y}$ and $x\in g^{-1}(y)$, then we have
 \begin{align*}
	 &\pmf{Y_{\vi}|X_{\neigh{\vi}}}(y_{\vi}|x_{\neigh{\vi}})\notag \\
	 &\stackrel{\textrm{(a)}}{=} \frac{\sum\limits_{x_{\vi}'\in
	 g_{\vi}^{-1}(y_{\vi})}\prod\limits_{C\in\sets{A}}\psi_C(x_{\vi}',x_{\not \vi}) }
	 {\sum\limits_{y_{\vi}'\in\sets{Y}_{\vi}}\sum\limits_{x_{\vi}''\in
	 g_{\vi}^{-1}(y'_{\vi})}\prod\limits_{C\in\sets{A}}\psi_C(x_{\vi}'',x_{\not \vi})}\\
	 &\stackrel{\textrm{(b)}}{=}
	 \frac{\prod\limits_{C\in\sets{A}\setminus\{C'(\vi)\}}U_C(y) \times \sum\limits_{x_{\vi}'\in g_{\vi}^{-1}(y_{\vi})}\psi_{C'(\vi)}(x_{\vi}',x_{\not \vi}) }
	  {\sum\limits_{y_{\vi}'\in\sets{Y}_{\vi}}\prod\limits_{C\in\sets{A}\setminus\{C'(\vi)\}}U_C(y'_{\vi},y_{\not \vi}) \times \sum\limits_{x_{\vi}''\in g_{\vi}^{-1}(y'_{\vi})}\psi_{C'(\vi)}(x_{\vi}'',x_{\not \vi})}\\
	  &\stackrel{\textrm{(c)}}{=} \frac{\prod\limits_{C\in\sets{A}}U_C(y) }
	  {\sum\limits_{y_{\vi}'\in\sets{Y}_{\vi}}\prod\limits_{C\in\sets{A}}U_C(y'_{\vi},y_{\not \vi})} \\
	  &= \pmf{Y_{\vi}|Y_{\neigh{\vi}}}(y_{\vi}|y_{\neigh{\vi}}),
 \end{align*}
 as desired. 
 In the above,
 (a) follows from \eqref{eq:pX-Ni} using
 $\pmf{Y_{\vi}|X_{\neigh{\vi}}}(y_{\vi}|x_{\neigh{\vi}}) = \sum_{x_{\vi}'\in
 g_{\vi}^{-1}(y_{\vi})} \pmf{X_{\vi}|X_{\neigh{\vi}}}(x_{\vi}'|x_{\neigh{\vi}})$ and the fact that
 the set of preimages $g^{-1}(y_{\vi})$, for $y_{\vi} \in \sets{Y}_{\vi}$, is a partition of
 $\sets{X}_{\vi}$; 
 (b) follows since at most one clique $C'(\vi)$ may strictly depend on $x_{\vi}$ where we define
 $U_{C}(g(x)):=\psi_C(x)$ for $C\in \sets{A}\setminus\{C'(\vi)\}$;
 and (c) follows since the potential
 function of $C'(\vi)$ depends on $x_{\not i}$ only via $y_{\not i}$
% due to the assumption in the proposition
 where we define
 $U_{C'(\vi)}(y):=\sum_{x_{\vi}'\in g_{\vi}^{-1}(y_{\vi})}\psi_{C'(\vi)}(x_{\vi}',x_{\not \vi})$, 
 see~\eqref{eq:projections} (or \eqref{eq:proof1:UC}).\qed

\subsection{Proof of Proposition~\ref{prop:infopreservation}}\label{proof:infopreservation}
Note that a lumping is information-preserving iff $H(X|Y) = 0$, then we have
% The proof of the necessary condition follows by $(a)$ the chain rule of entropy and $(b)$ the fact that conditioning reduces entropy. Indeed,
 \begin{align*}
	 0 = \ent{X|Y} &\stackrel{\textrm{(a)}}{=} \sum_{\vi=1}^{|\sets{V}|} \ent{X_{\vi}|X_1^{\vi-1},Y}\\
	 &\stackrel{\textrm{(b)}}{\ge} \sum_{\vi=1}^{|\sets{V}|} \ent{X_{\vi}|X_{\not \vi},Y}\\
	 &\stackrel{\textrm{(c)}}{=} \sum_{\vi=1}^{|\sets{V}|} \ent{X_{\vi}|X_{\not \vi},Y_{\vi}}\\
	 &\stackrel{\textrm{(d)}}{=} \sum_{\vi=1}^{|\sets{V}|} \ent{X_{\vi}|X_{\neigh{\vi}},Y_{\vi}},
 \end{align*}
 where (a) is by the chain rule of entropy, (b) by the fact that conditioning reduces
 entropy, (c) by $Y_{\vj}=g_{\vj}(X_{\vj})$ for $\vj\neq \vi$, and (d) by the fact that $X$ is a $\graph$-MRF.

The proof of the sufficient condition for chordal graphs makes use of the following lemma stating that for a MRF on a chordal graph the entropy can be decomposed in a particular form.
\begin{lem}\label{lem:chordal}
If $\graph$ is chordal, then there exists a permutation $\perm_1,\dots,\perm_N$ of the vertex indices such that
 \begin{equation}\label{eq:ITEq}
  \ent{X} = \sum_{i=1}^N \ent{X_{\perm_i}|X_{A_{\perm_i}}}
 \end{equation}
 for $A_{\perm_i}=\neigh{\perm_i}\cap \{\perm_1,\dots,\perm_{i-1}\}$.  
\end{lem}

%Thus, with this lemma in (a) and the fact that conditioning reduces entropy in (b), the proof is completed with
%\begin{align*}
% \ent{X|Y} &\stackrel{\textrm{(a)}}{=} \sum_{i=1}^N \ent{X_{\perm_i}|X_{A_{\perm_i}},Y}\\
% &\stackrel{\textrm{(b)}}{\le}\sum_{i=1}^N \ent{X_{\perm_i}|X_{A_{\perm_i}},Y_{\perm_i}}.
%\end{align*}
%In (a), we additionally require that $X$ is a $\graph$-MRF when conditioned on realizations of $Y$. The proof of this result follows along similar lines as the corresponding result for Markov chains in~\cite[Sec.~3.3]{GeigerTemmel_kLump}. \qed

%{\color{blue}
%First, note that, conditioned on $Y$, the RV $X$ is still a $\graph$-MRF. (This follows
%since the RVs $Y_{\vi}, \vi \in \sets V$ correspond to leaf nodes in the graph $\graph_{X,Y}$, see
%~\cite[Sec.~3.3]{GeigerTemmel_kLump} for details.)
%Hence, as desired, we have 
%\begin{align*}
% \ent{X|Y} &\stackrel{\textrm{(a)}}{=} \sum_{i=1}^N \ent{X_{\perm_i}|X_{A_{\perm_i}},Y}\\
% &\stackrel{\textrm{(b)}}{\le}\sum_{i=1}^N \ent{X_{\perm_i}|X_{A_{\perm_i}},Y_{\perm_i}},
%\end{align*}
%where (a) is by the above lemma and (b) by the fact that conditioning reduces entropy. \qed
%}

%{\color{magenta}
To prove the sufficient condition, we have
\begin{align*}
%	&H(X|Y) + H(Y) = H(X,Y)  \\
%	& \stackrel{(a)}{=} H(X)  \\
	H(X) 
	& \stackrel{\textrm{(a)}}{=} \sum_{i=1}^N \ent{X_{\perm_i}|X_{A_{\perm_i}}} \\
	& \stackrel{\textrm{(b)}}{=} \sum_{i=1}^N \ent{X_{\perm_i}, Y_{\perm_i}|X_{A_{\perm_i}}} \\
	& \stackrel{}{=} \sum_{i=1}^N \big( \ent{Y_{\perm_i}|X_{A_{\perm_i}}} + \ent{X_{\perm_i}|X_{A_{\perm_i}},Y_{\perm_i}} \big)  \\
	& \stackrel{\textrm{(c)}}{=} \sum_{i=1}^N \big( \ent{X_{\perm_i}|X_{A_{\perm_i}},Y_{\perm_i}} + \ent{Y_{\perm_i}|X_{A_{\perm_i}}, Y_{\perm_1, \dots, \perm_{i-1}}} \big) \\
	& \stackrel{\textrm(d)}{\leq} \sum_{i=1}^N \big( \ent{X_{\perm_i}|X_{A_{\perm_i}},Y_{\perm_i}} + \ent{Y_{\perm_i}| Y_{\perm_1, \dots, \perm_{i-1}}} \big) \\
	& \stackrel{\textrm(e)}{=} \sum_{i=1}^N \ent{X_{\perm_i}|X_{A_{\perm_i}},Y_{\perm_i}} + \ent{Y},
\end{align*}
where (a) is by Lemma~\ref{lem:chordal}, (b) is because $Y_{\perm_i}$ is a function of $X_{\perm_i}$,
(c) is because $Y_{\perm_i}$ and $Y_{\{\perm_1, \dots, \perm_{i-1}\}}$ are independent conditioned on
$X_{A_{\perm_i}}$, (d) is because conditioning reduces the entropy, and (e) is by the entropy chain rule.
Now the claim follows from $H(X|Y) = H(X) - H(Y)$, which is true since $Y$ is a function of
$X$. \qed
%}

\begin{IEEEproof}[Proof of Lemma~\ref{lem:chordal}]
 	A maximum cardinality search \cite[Section~3.2.4]{pearl-reasoning1988} provides the desired
	permutation, where the permutation can also be viewed as an orientation of the graph into a
	directed acyclic graph (DAG). Since the original graph is chordal, the resulting DAG is such that
	the tails of any two converging arrows are adjacent, and so d-separation (on the DAG) and vertex
	cuts (on the original graph) are equivalent. That is, they describe the same independence
	structure, i.e., the same family of distributions. Now the factorization form~\eqref{eq:ITEq} is easy
	to obtain from the DAG.
\end{IEEEproof}

%\section*{Acknowledgments}
%The authors wish to thank Raymond W. Yeung for valuable discussions.
%The work of Bernhard C. Geiger has been supported by the HiDALGO project and has been partly funded
%by the European Commission’s ICT activity of the H2020 Programme under grant agreement number
%824115.

\balance

\fi

\section*{Acknowledgments}
The authors wish to thank Raymond W. Yeung for valuable discussions and for pointing out~\cite{Sadeghi_Marginalization}.
The work of Bernhard C. Geiger has been supported by the
HiDALGO project and has been partly funded by the European Commission’s ICT activity of the H2020
Programme under grant agreement number 824115. 
The Know-Center is funded within the
Austrian COMET Program - Competence Centers for Excellent Technologies - under the auspices of the
Austrian Federal Ministry of Climate Action, Environment, Energy, Mobility, Innovation and
Technology, the Austrian Federal Ministry of Digital and Economic Affairs, and by the State of
Styria. COMET is managed by the Austrian Research Promotion Agency FFG.

\bibliographystyle{IEEEtran}
\bibliography{IEEEabrv,references}

\end{document}